\documentclass[twoside,11pt]{article}

\usepackage{jmlr2e}
\usepackage{algorithm}
\usepackage{url}
\usepackage{algorithmic}
\usepackage{amsmath}
\usepackage{amssymb}
\usepackage{graphicx}
\usepackage{listings}
\usepackage{color}
\usepackage{xcolor}
\usepackage{hyperref}

\newcommand{\ie}{\emph{i.e.}}
\newcommand{\eg}{\emph{e.g.}}

\newcommand{\vct}[1]{\ensuremath{\mathbf{#1}}}

\newcommand{\adversarialib}{\texttt{AdversariaLib}}
\newcommand{\repository}{\url{http://sourceforge.net/projects/adversarialib}}
\newcommand{\website}{\url{http://comsec.diee.unica.it/adversarialib}}

\newcommand{\knowref}[2][]{\ifx&#1&({\it k.\ref{#2}})\else({\it k.\ref{#1}-\ref{#2}})\fi}
\newcommand{\capref}[2][]{\ifx&#1&({\it c.\ref{#2}})\else({\it c.\ref{#1}-\ref{#2}})\fi}

\firstpageno{1}

\begin{document}

\title{\adversarialib{}: An Open-source Library for the Security Evaluation of Machine Learning Algorithms Under Attack}

\author{\name Igino Corona \\
\name Battista Biggio \\
\name Davide Maiorca \\
       \addr Dept. of Electrical and Electronic Engineering\\
       University of Cagliari\\
       Piazza d'Armi, 09123, Cagliari, Italy\\
       \email \{igino.corona,battista.biggio,davide.maiorca\}@diee.unica.it}

\editor{}

%

\maketitle

\begin{abstract}
We present \adversarialib{}, an open-source python library for the security evaluation of machine learning (ML) against carefully-targeted attacks. It supports the implementation of several attacks proposed thus far in the literature of adversarial learning, allows for the evaluation of a wide range of ML algorithms, runs on multiple platforms, and has multi-processing enabled.
The library has a modular architecture that makes it easy to use and to extend by implementing novel attacks and countermeasures.
It relies on other widely-used open-source ML libraries, including \emph{scikit-learn} and \emph{FANN}. Classification algorithms are implemented and optimized in C/C++, allowing for a fast evaluation of the simulated attacks. The package is distributed under the GNU General Public License v3, and it is available for download at \repository{}.
\end{abstract}


\section{Introduction} 
Machine learning algorithms have been increasingly adopted in security applications like spam, intrusion and malware detection, mainly due to their ability to generalize, \ie, to potentially detect novel attacks or variants of known ones. However, in adversarial settings such as the aforementioned ones, the underlying assumption of data \emph{stationarity}, \ie, that both training and test data are drawn from the same (though possibly unknown) distribution, is often violated by intelligent and adaptive adversaries that purposely manipulate data to compromise learning. This has recently raised the issue of evaluating the security of learning algorithms to carefully-targeted attacks, besides designing suitable countermeasures, as pointed out in the growing research area of \emph{adversarial machine learning}~\citep{barreno06-asiaccs,huang11,biggio14-tkde}. 

In this work we present \adversarialib{}, an open-source python library that can be exploited to this end. To the best of our knowledge, this is the first open-source library that implements the process of security evaluation recently advocated by \cite{biggio14-tkde}. 
The idea behind a \emph{security evaluation} is to \emph{proactively} anticipate the behavior of the adversary to identify potential vulnerabilities of machine learning algorithms, and to design suitable countermeasures, \emph{before} the corresponding attacks may actually occur.  To this end, potential attacks against the given classifier are \emph{simulated} according to a given adversary's model through an ad hoc manipulation of training and test data, and then their impact on the targeted classifier's performance is evaluated. The core of this procedure is clearly the way in which the training and the test data are modified, depending on the considered attack.
For instance, in the \emph{evasion} setting, the adversary can only manipulate malicious samples (\eg{}, spam emails) at test time to evade detection, while training data remains unchanged. How each sample is manipulated depends on specific assumptions made on the adversary's model, \ie{}, on her goal, knowledge of the attacked system, and capabilities of manipulating the data \citep{biggio14-tkde,biggio13-ecml}. It is thus clear that the attack strongly depends on the targeted classifier. In practice, however, only rarely the adversary may  know the targeted classifier completely: although she may realistically know the kind of algorithm used, she may not know the parameters learned after training (\eg, the weights assigned to each feature by a linear classifier). 


Another interesting feature of \adversarialib{} is indeed the possibility of learning an estimate of the targeted classifier, which we refer to as \emph{surrogate} classifier. This assumes that the adversary can collect a set of surrogate training data, ideally drawn from the same distribution as the targeted classifier's training set, retrieve the labels assigned to them by the targeted classifier, and learn a \emph{surrogate} classifier on such data. In \cite{biggio13-ecml} we showed that attacking a surrogate classifier learned on a relatively small training set may lead to evade the targeted classifier with high probability. This is an example of how machine learning can be offensively exploited to defeat machine learning-based defenses. \adversarialib{} natively implements this mechanism to simulate \emph{realistic} attacks in which the adversary has limited knowledge of the attacked system.

We present the library's architecture and implementation in Sect.~\ref{sect:architecture}, as well as an attack example in Sect.~\ref{sect:attack-example}. Conclusions and future extensions are discussed in Sect.~\ref{sect:conclusions}.

\section{\texttt{AdversariaLib}: architecture and implementation} 
\label{sect:architecture}

\begin{figure}[htbp]
\begin{center}
\includegraphics[width=0.99\textwidth,trim=40 286 25 20, clip]{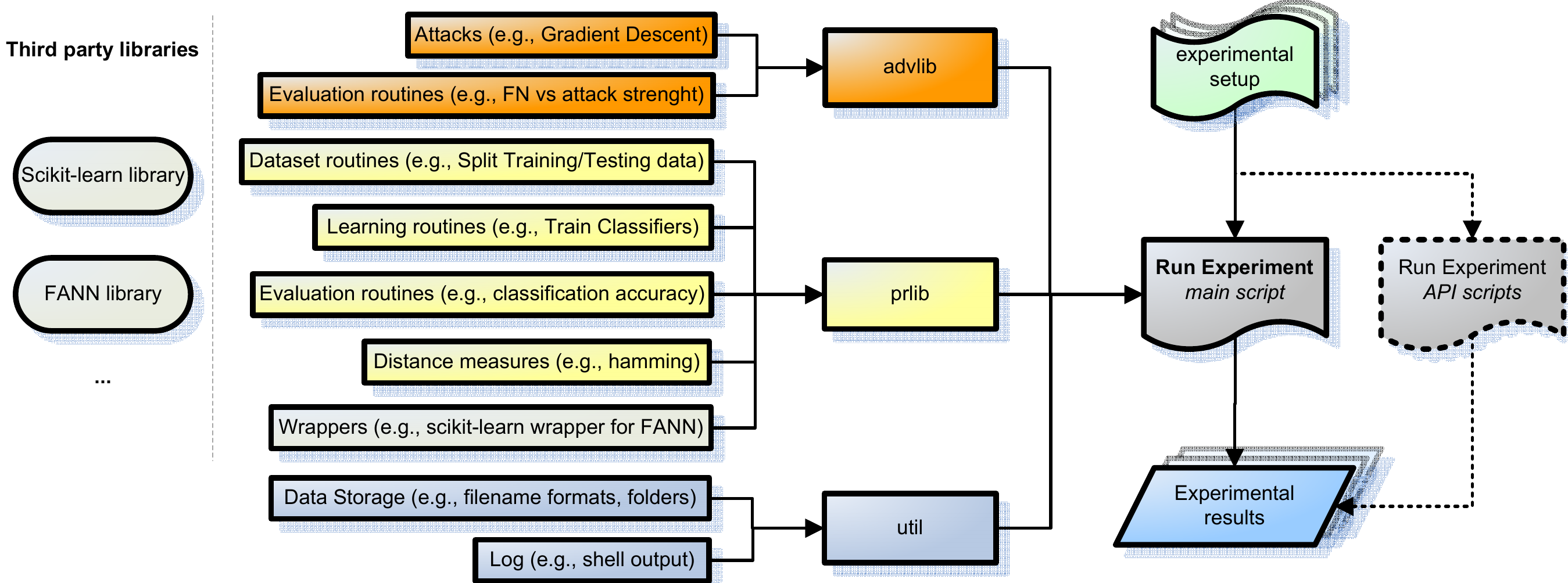}
\caption{Architecture and main components of \adversarialib{}.}
\label{fig:architecture}
\end{center}
\end{figure}

\adversarialib{}, as shown in Fig.~\ref{fig:architecture}, is structured into three main modules. The first, \verb=advlib=, implements the attack algorithms and functions for the evaluation of classifiers under attack. The second, \verb=prlib=, implements standard functions for handling datasets, defining distances between samples in feature space, learning classifiers, and evaluating their performance. To this end, \verb=prlib= implements suitable wrappers that exploit other open-source libraries, such as \emph{scikit-learn}~\citep{scikit-learn} and \emph{FANN}.\footnote{\emph{FANN} is an open-source C library for learning Fast Artificial Neural Networks (not natively supported by \emph{scikit-learn}), available at \url{http://leenissen.dk/fann/wp/}.}
The third, \verb=util=, contains library functions for storing datasets, classifiers, and results in the filesystem, and for logging the progress of current experiments (on the standard output).

The library's main script, named \texttt{runexp}, implements a rather general experimental scheme that can be used to evaluate the security of different machine learning algorithms against several kinds of adversarial attacks. 
In particular, it executes the following steps: $(i)$ randomly split data into training-test pairs (if such pairs are not given initially); $(ii)$ for each training portion, learn the corresponding classifier(s);  $(iii)$ attack each learned classifier; and $(iv)$ store the results.
Any experiment is executed according to the parameters specified in the \emph{setup} file, such as, \eg, number of training-test pairs, type and parameters of each targeted classifier, and type and parameters of the chosen attack(s).
It is possible to run different experiments through the proper definition of different setup files. Experimental results are saved according to the format defined in the \verb=util= module, and depending on the given attack. \adversarialib{} also provides a set of \emph{API} scripts to run simple experiments, such as training a classifier, attacking it, and evaluating its classification accuracy.

\textbf{Implementation}. \adversarialib{}, as well as its dependencies (third-party libraries) is fully open-source, multi-platform, and can be used flawlessly in all major operating systems, since it is written mainly in Python. An important aspect is that Python allows researchers to easily and quickly implement new functionalities, and has a huge developers community. Moreover, an in-depth code optimization can be easily achieved exploiting C/C++ language bindings for Python, as performed by the external libraries \emph{scikit-learn} and \emph{FANN}.
We carefully documented the whole library, with an extensive set of usage examples, as described in the official website \website{}. 
We also provided a detailed documentation on how to implement and add new attacks to the library, to make it easy to extend. Each attack implementation must provide a simple, general interface, and the related files must be put in a separate folder within the \verb=advlib= module (see Fig.~\ref{fig:architecture}). 

\textbf{FANN and Matlab wrappers}. We implemented a scikit-learn wrapper for FANN classifiers (see \texttt{prlib/classifier/mlp.py}). 
Following the same approach, novel classifiers, \eg, learning algorithms robust to adversarial attacks, can be easily added to the library. 
%
We also provide an open-source wrapper that allows us to configure and run experiments using \adversarialib{} directly from the Matlab environment. In addition, the corresponding results can be exported as \texttt{PDF} files. This wrapper relies on the APIs of \adversarialib{}, and can thus easily execute the evasion attack described by~\citet{biggio13-ecml}. An example is presented in the next section. This wrapper is freely downloadable from the official \adversarialib{} repository.

\section{Attack example on handwritten digit recognition}
\label{sect:attack-example}

\begin{figure}[t]
\begin{center}
\includegraphics[width=0.75\textwidth,trim=0 24 0 0, clip]{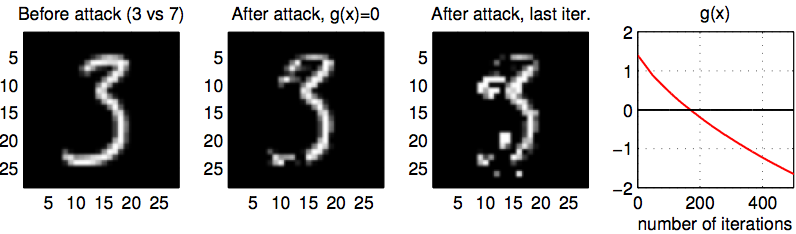}
\caption{Gradient-based evasion of a linear SVM trained on handwritten digits.}
\label{fig:digits}
\end{center}
\end{figure}

We report here a simple example to visually demonstrate the effectiveness of a gradient-based evasion attack, adapted from our recent work in \cite{biggio13-ecml}.
This example can be run using the Matlab wrapper of \adversarialib{} and, in particular, the script \verb=main_mnist.m=. 
A linear SVM is first trained to discriminate between two classes of handwritten digits from the MNIST dataset~\citep{LeCun95}, \ie, \texttt{3} (positive class) and \texttt{7} (negative class), and then, the former is manipulated to evade detection. The attack iteratively modifies the initial \texttt{3} to minimize the SVM's discriminant function $g(\vct x)$ through gradient descent.
Results are shown in Fig.~\ref{fig:digits}. The leftmost \texttt{3} represents the initial, non-manipulated attack sample. The second and third depicted digits represent \emph{evasion points}, \ie, manipulated \texttt{3}s that are misclassified as \texttt{7}, respectively as soon as the SVM's discriminant function $g(\vct x)$ becomes non-positive (\ie, the first found evasion point during the descent), and after more iterations.
This example not only highlights that few manipulations are required to the initial \texttt{3} to be misclassified as \texttt{7}, but also that they are not even required to mimic the targeted class (the manipulated \texttt{3}s evade detection without resembling a \texttt{7} at all). This confirms the vulnerability of standard learning algorithms to well-crafted attacks, and the need for more \emph{secure} learning algorithms, as suggested by \cite{barreno06-asiaccs,huang11,biggio14-tkde}.

\section{Conclusions and future work}
\label{sect:conclusions}
\adversarialib{} is the first open-source library for the security evaluation of machine learning against carefully targeted-attacks. Currently, it implements the gradient-based evasion attack described in \cite{biggio13-ecml}. As the library is modular and easy to extend, novel and more sophisticated attacks can be easily implemented, like \emph{poisoning}~\citep{biggio12-icml,kloft10}, as well as countermeasures and secure classifiers, hopefully, with the help of an emerging community of users and developers.



\end{document}